# Terahertz Phonons coherently couple Ultrashort Solitons inside a Cavity

Alexandra Völkel[1] and Georg Herink[1*]

[1] Experimental Physics VIII – Ultrafast Dynamics, University of Bayreuth, Germany

**Ultrafast atomic vibrations mediate heat transport, serve as fingerprints for chemical bonds and drive phase transitions in condensed matter systems[1–3]. Light pulses shorter than the atomic oscillation period can not only probe, but even stimulate and control collective excitations[4–7]. In general, such interactions are performed with free-propagating pulses. Here, we demonstrate intra-cavity excitation and time-domain sampling of coherent optical phonons inside an active laser oscillator. Employing real-time spectral interferometry, we reveal that Terahertz beats of Raman-active optical phonons can bind temporal solitons to bound-states – termed "Soliton molecules" – and we resolve a coherent coupling mechanism of phonon and intra-cavity soliton motion[8]. Concurring electronic and nuclear nonlinearities generate distinct soliton trajectories and, effectively, enhance the coherent Raman signal. As a result, we observe that the inter-soliton motion automatically performs highspeed time-domain Raman spectroscopy of the gain medium. In particular, our results pinpoint the impact of Raman-induced soliton interactions in crystalline laser media and microresonators[9,10], and offer unique perspectives towards ultrafast nonlinear phononics by exploiting the enhanced coupling of atomic motion and solitons inside a cavity[11,12].**

Time-resolved sampling of coherent excitations comprises a key concept of pump-probe spectroscopy in experimental quantum physics and ultrafast sciences[13]. The probing of atomic motion on its natural timescale unveils phonon lifetimes, vibrational dephasing and couplings in condensed matter[1,4,7,14]. Particularly in transparent media, ultrashort pump pulses can impulsively stimulate coherent phonon motion via Raman scattering non-resonant with the electronic system. This impulsive Raman excitation allows for preparing, manipulating and controlling coherent phonon states and it can occur at every roundtrip inside an active ultrafast laser cavity. The dynamics commonly remain hidden due to phonon lifetimes that are short compared to typical cavity roundtrip periods. Here, we study inter-soliton motion and unravel the link between impulsive stimulated Raman excitation and the interaction of ultrashort temporal solitons, and we identify Raman-active optical phonons as the key constituent for soliton binding[15–17] and complex inter-soliton trajectories[8]. Moreover, we demonstrate that real-time optical interferometry of soliton interactions inside a laser cavity may serve as an unprecedented tool for coherent tracking of phonon dynamics and stimulated Raman spectroscopy (SRS) in the temporal domain at high speed[3].

Recently, real-time optical spectroscopy based on the time-stretch dispersive Fourier transform (TS-DFT) has enabled high throughput measurements in laser science and nonlinear optics, yielding consecutive single-shot spectra exceeding 100 million frames per second. Following on observations in passive nonlinear systems[18,19], the scheme has been applied to nonlinear behavior inside ultrafast oscillators, opening up views into the buildup of soliton mode-locking and complex multi-soliton interactions in real-time[20,21,8]. In particular, temporal solitons are observed to form stable and meta-stable bound states - termed "soliton molecules" - with non-commensurate temporal separation distances, spanning from picoseconds down to few 10-femtoseconds. Deterministic switching between such soliton molecules upon external control was also demonstrated[22]. The temporal soliton separations in these systems match various earlier reports on double-pulsing in Kerr-lens modelocked Ti:sapphire lasers – the underlying mechanisms, however, have remained elusive[15–17].

---

* Corresponding author: georg.herink@uni-bayreuth.de

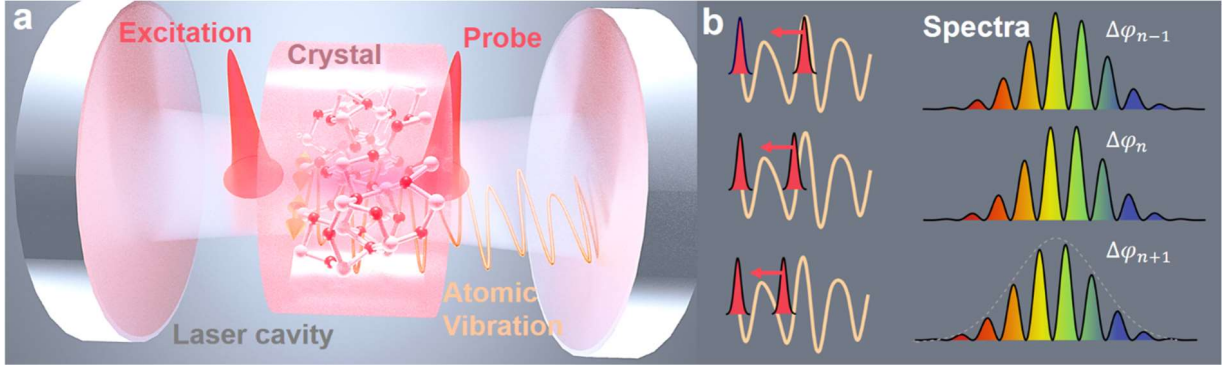

**Figure 1: Time-domain Soliton Raman Sampling inside a Cavity: a,** Two ultrashort solitons (red) circulate inside an active laser cavity: The leading soliton excites a coherent lattice vibration (orange) in the gain crystal, the trailing pulse samples the refractive index modulation. **b,** Upon their approach, the trailing soliton traces the temporal waveform of the lattice vibration via its relative phase (left). The relative phases $\Delta\varphi$ of consecutive single-shot spectra encode the temporal Raman response (right).

In this work, we analyze real-time measurements of transient dynamics in Kerr-lens modelocked laser oscillators, based on experimental data and previous reports[8,16]. According to a common design of mode-locked Titanium-doped sapphire oscillators, the crystal serves as a broadband gain medium and as an ultrafast intensity-dependent amplitude modulator, enabling passive mode-locking via nonlinear focusing due to the Kerr effect[23,24]. Upon tuning of the resonator stability for single-pulse operation, mode-locking of two solitary pulses is obtained, which circulate almost independently inside the resonator. Yet, interactions based on the laser gain saturation and nonlinear refraction induce two different group velocities, and thus, generate relative motion of both pulses[25]. As a result, the solitons can autonomously sweep through their temporal separation and effectively perform degenerate pump-probe scanning in an all-optical and autonomous fashion. For large temporal separations of several hundred picoseconds, such relative motion is perceivable via fast real-time photodetection of the output intensity.

Pico- to femtosecond separations are not resolved via direct photodetection, but can be tracked on a continuous, single-shot basis employing TS-DFT and spectral interference. Solitons form meta-stable bound-states at separations in the picosecond to few 10-fs range, observable via field-autocorrelations from real-time spectral interferograms[8]. The states are incommensurably spaced and exhibit varying degrees of stability. We extract these separations, displayed as maxima in the cumulative Fourier transforms of consecutive spectra in Figure 2b, and they coincidence with static bound-states reported in early studies of Kerr-lens mode-locking in Ti:Sapphire, as indicated by red marks in Fig. 2b. By this means, we can directly unravel the underlying binding mechanism of frequently observed bound-state separations: we relate the data to the temporal Raman response function of the host medium, shown in Fig. 2b (orange line). The active Titanium-doped sapphire (Ti:Al$_2$O$_3$) medium corresponds to the crystal structure of the rhombohedral space group R-3c. The birefringent crystal is aligned with the optical axis parallel to the linear polarization of the beam and within the plane of incidence, and the a-facet aligned to Brewster's angle. Referring to Porto's notation, the Raman selection rules for the corresponding *X(ZZ)X*-geometry allow for accessing symmetric A$_{1g}$-modes, given by two dominant optical phonons at 645 1/cm and 418 1/cm wavenumbers, or frequencies of 16 THz and 13 THz, respectively[26,27]. Following impulsive excitation by the first soliton at delay $\tau = 0$, the two coherent modes generate an exponentially decaying waveform and the resultant local maxima of phonon beats coincide with the bound-state separations. The physical mechanisms will be discussed in more detail below.

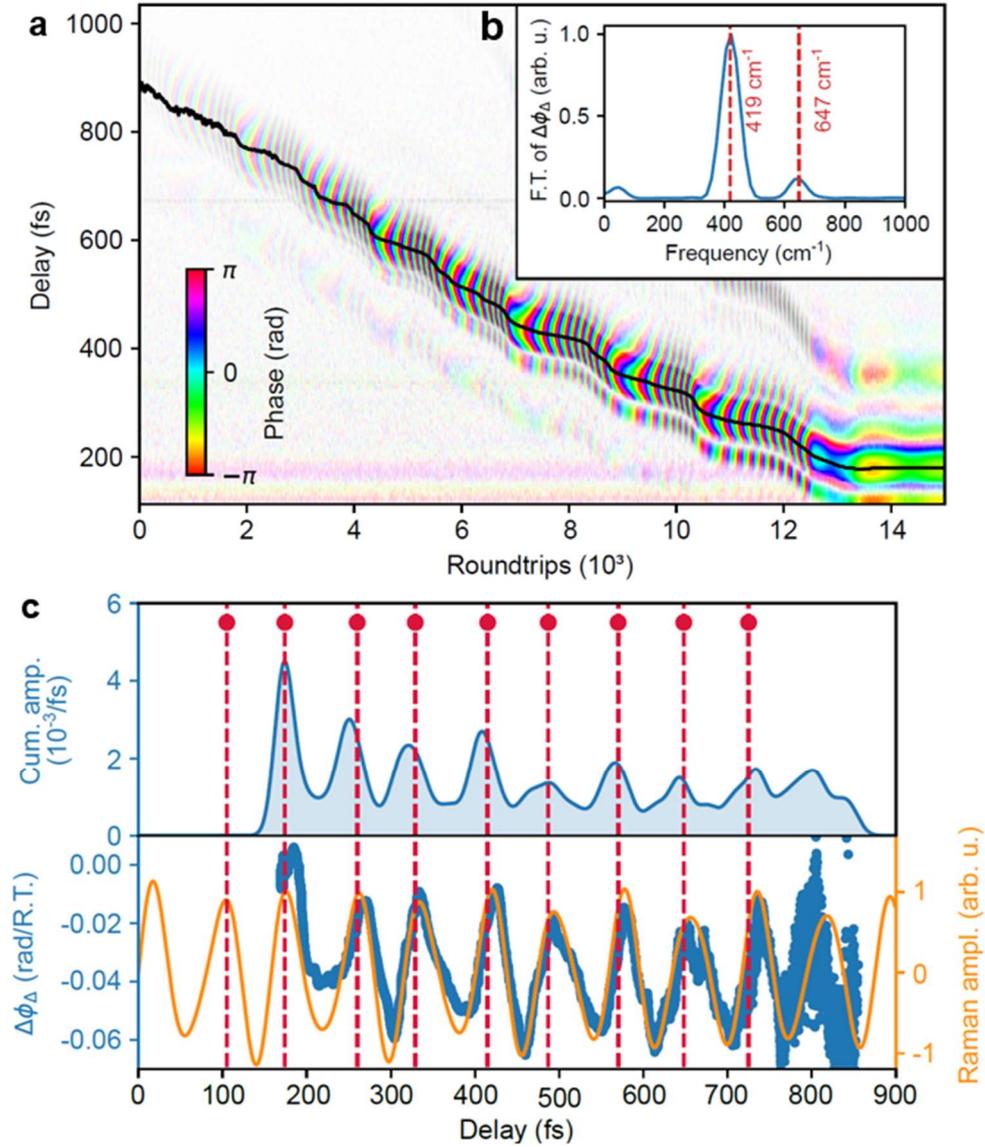

**Figure 2: Transient soliton approach and extraction of the temporal Raman response. a,** The spectral phase evolution in the interferograms of two approaching solitons features a characteristic pattern that encodes the Raman response. The phase evolution (color coded) as a function of cavity roundtrips is extracted at the soliton separations (black line), obtained from the maxima of the corresponding Fourier amplitude (intensity weighted by Fourier amplitude). Experimental data from Ref. [8]. **b,** The resulting Raman spectrum obtained from Fourier Transformation of the measured signal $\Delta\phi_\Delta$ (see c) and corresponding Raman peaks from literature [26]. **c,** Top: The cumulative sum of all Fourier amplitudes as a function of soliton separation exhibits peaks at meta-stable separations. The latter coincide with bound-states observed by Kitano et al. (Ref [16]), marked by red dots and dashed lines. Bottom: The relative phase-change per roundtrip, extracted via consecutive double-referencing, yields the full temporal phonon amplitude waveform, as evidenced by the theoretical Raman response (orange line). Soliton motion persists at phonon beats of the lattice vibration.

Furthermore, our analysis allows for tracking the continuous vibrational phonon waveform of Raman-active modes in the time-domain based on a single cavity soliton trajectory, captured via a real-time spectrogram within an acquisition time below 170 μs. For data extraction, we focus on the evolution of the relative spectral phase. Specifically, we introduce "consecutive double-referencing": whereas the spectral phase $\Delta\varphi(\omega)$ of a spectral interferogram of two pulses with fields $E_{1,2}$,

$$|E(\omega)|^2 = \frac{1}{2}\{|E_1(\omega)|^2 + |E_2(\omega)|^2\} + |E_1(\omega)| \cdot |E_2(\omega)| \cdot \cos(\omega_0 \tau + \Delta\varphi(\omega)),$$

evaluated at roundtrip $n$ encodes the relative temporal phase $\Delta\varphi_n(\tau)$ of a bound-state, the difference to the previous interferogram tracks the relative phase *change* for a single cavity roundtrip $\Delta\varphi_\Delta(\tau) = \Delta\varphi_n - \Delta\varphi_{n-1}$, sketched in Fig. 1b. This difference signal $\Delta\varphi_\Delta(\tau)$ is immune to arbitrary prevailing intensity differences within the soliton pair due to the soliton generation process, and it is stationary at a given soliton separation $\tau$. However, this observable is sensitive to round-trip gain differences due to the coupling with the Raman modes, sampled at varying inter-soliton separations $\tau$. We map this double-referenced signal as a function of effective pump-probe delay $\tau$ and find that this observable tracks the Raman response with high fidelity, evidenced by the theoretical Raman response constructed from referenced Raman data [26]. Hence, our approach yields complex-valued Raman spectra via Fourier-Transformation[28]. The obtained amplitude spectrum is shown in Figure 2c. We note that further optimization with respect to signal-to-noise, i.e. data averaging, or sampling speed has not been applied at this point. Moreover, the quality of the Raman signal depends on the actual cavity alignment and a divers set of trajectories is obtained for different operational parameters (see Supplementary Information). Also, deviations from a purely linear motion of soliton trajectories may result in deviations in the sampled Raman waveform (e.g. towards the end of motion at $\tau \sim 200 fs$).

Generally, the intrinsic data quality results from the combination of several critical requirements for coherent Raman spectroscopy: First, intra-cavity enhancement and dispersion compensation ensure intense ultrashort pump/probe pulses at high repetition rate; second, spectral filtering via the laser gain confines both pump and probe spectra, thus, allowing for sustained spectral coherence; and third, the Kerr-lensing effect enhances the effective Raman response via the coupling to delay-dependent gain and loss. This spatial Raman effect differs from Raman-induced Kerr-effect (RIKE) schemes measuring polarization modulation[29], and this Raman Kerr-lensing has recently been demonstrated for extra-cavity SRS[30]. In contrast to predominate SRS implementations, including dual-frequency comb spectroscopy[31] or TS-DFT for spectral-domain Raman detection[32,33], the Raman information in the presented intra-cavity scheme is encoded in the relative phase instead of pump or probe beam intensities[34,35]. This approach may offer both intrinsic signal-enhancement and accelerated acquisition speed[3], yet potential operational modes, fundamental limitations and practical designs for reliable high-speed spectroscopy require further in-depth analysis.

We now discuss the physical mechanism of the cavity soliton interaction inside a modelocked resonator via Raman phonons and present results from a numerically evaluated master-equation model based on the complex Ginzburg–Landau equation[36]. We distinctly reproduce the experimental observations, including characteristic soliton trajectories in spite of various simplifications (see Supplementary Information for details). Generally, the potential interaction regimes are highly divers and may sensitively depend on various cavity perturbations. Still, we were able to pinpoint and reproduce the fundamental sequence which enables intracavity Raman sampling in both experiment and theory.

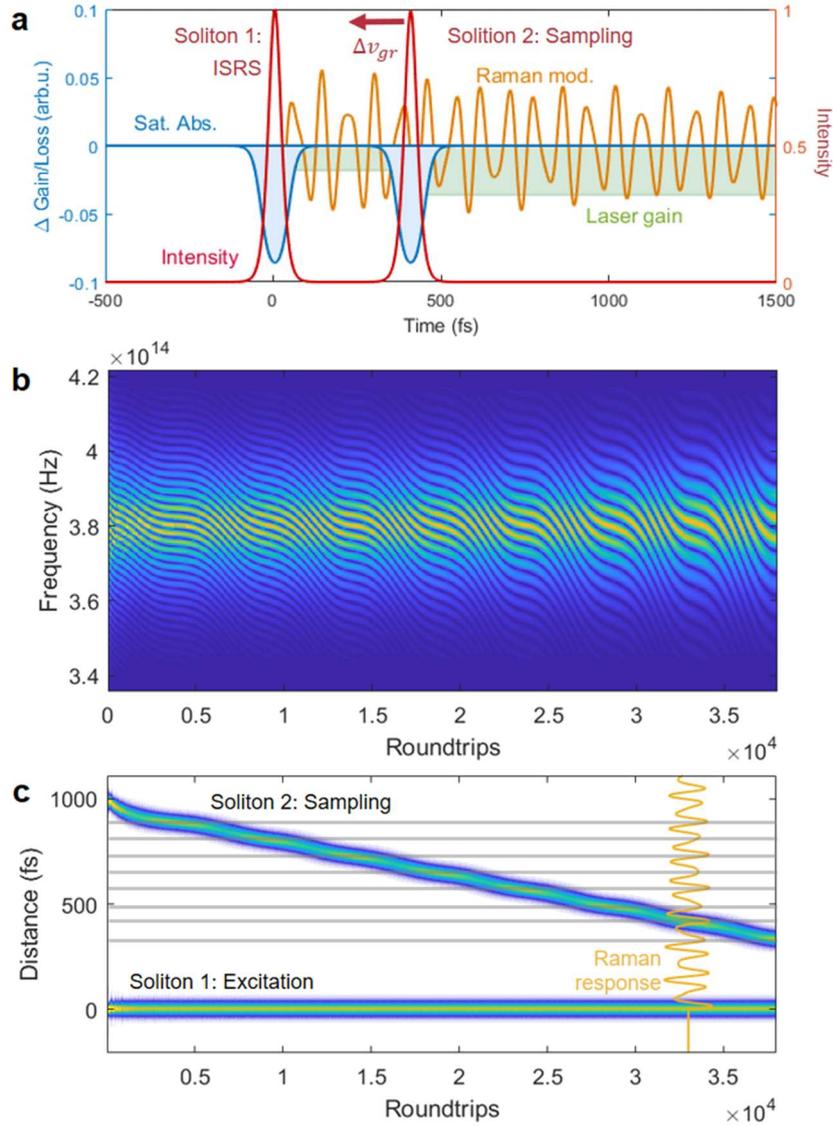

**Figure 3: Simulations of soliton interactions: a,** Intra-cavity Soliton Raman Sampling is based on the relative motion of two solitons inside an ultrafast laser provided by group velocity differences $\Delta v_{gr}$ due to gain depletion and self-steepening. Soliton 1 excites coherent lattice vibrations via impulsive simulated Raman scattering. The second soliton samples the Raman waveform as ultrafast modulations of the relative phase – arising from the combined nonlinear electronic and nuclear gain modulation in the Kerr-lens configuration. **b,** Simulated evolution of single-shot spectral interferograms within in the laboratory timeframe, i.e., as a function of cavity roundtrips. The stepwise soliton approach manifest via increasing fringe periods, and the spectral phase evolution encodes the Raman response (Intensity: color-coded, blue to yellow). **c,** Corresponding intra-cavity intensities in the co-moving timeframe of the leading soliton: The trailing soliton samples the Raman response (inset, orange line) and the relative motion halts at maxima of the Raman response - generating metastable separations at the beating maxima of both Raman modes (indicated as grey lines).

According to our model, the interaction process evolves via the following coupling sequence, sketched in Fig. 3a: Initially, two solitons are formed inequidistantly within the resonator and experience different laser gain due to transient gain depletion (green). The intricate balance of dissipative cavity solitons[37] allows for solutions at different intensities, however, the electronic $\chi^2_{el}$-nonlinearity induces a group velocity difference $\Delta v_{gr}$ via self-steepening, effectively slowing the leading pulse. In Raman-active media, each femtosecond soliton impulsively stimulates the collective coherent lattice vibration modes that are accessible upon nonlinear electronic mixing within the broad spectral bandwidth[5]. The collective lattice motion (orange) excited by the leading pulse contributes a resonant nuclear component to the effective nonlinear refractive index modulation $\Delta n$ at delay $\tau$, $\Delta n(t-\tau) = n_{2,el}I(t-\tau) + \int_0^\infty R(u)I(t-\tau-u)\,du$, with the response function $R(u)$. In the impulsive limit, the delay-dependent nuclear index modulation experienced by the trailing pulse yields: $\Delta n_{2,nuc}(\tau) = R(\tau) \cdot I$. The effective Raman refractive response for the sapphire laser crystal after impulsive stimulation corresponds to a free exponential decay of in this case $n = 2$ dominant optical phonon modes[38]: $R(\tau) = \sum_{i=1}^n R_n \cdot \exp(-\tau/T_{R_n}) \cdot \sin(2\pi\,\Omega_n\tau)$ at frequencies $\Omega_1 = 16\,THz$ and $\Omega_2 = 13\,THz$, respectively. The Raman contribution presents a subtle perturbation to the nonlinear index modulation, which is dominated by the electronic nonlinearity, and commonly remains indiscernible in this experimental probing configuration. However, inside the mode-locked cavity, the additional nonlinear index enhances the *spatial* electronic Kerr-lensing effect[23] and, thus, generates increased laser gain at phases of the coherent lattice vibration with positive nuclear index contribution. Vice versa, increased gain of the trailing soliton manifests as a relative phase change per roundtrip mediated by the electronic self-phase modulation – and, thus, provides enhanced detection sensitivity observable via consecutive double-referenced detection. Effectively, the coupling of Kerr nonlinearities with roundtrip gain amplifies the Raman response and renders the waveform detectable with realtime single-shot spectroscopy.

Moreover, the Raman excitation modulates the intensity of the second soliton and reduces the group velocity mismatch within the soliton pair at local maxima of the Raman waveform. Depending on the balance of gain depletion and lattice excitation strengths, the relative motion decelerates - apparent as meta-stable bound-states - but may possibly not stop: Upon further approach, the nuclear index modulation changes sign and reduces the effective index: Vice versa, the intensity of the trailing pulse decreases and relative soliton motion accelerates. Upon successive approach of the solitons towards zero delay, the phonon amplitudes experienced by the trailing pulse increase until the Raman interaction eventually balances initial intensity differences of both solitons and a phase-stable bound-state establishes.

The overall sequence temporally traces the Raman response, and we reproduce the soliton trajectory in our simulations: The resultant interferometric spectrogram in the laboratory time-frame, i.e. as a function cavity roundtrips, is displayed in Figure 3b. The roundtrip intensity in the co-moving time-frame (Fig. 3c) reveals characteristic alternating periods between accelerated and decelerated relative soliton motion. The metastable bound-states correspond to local beating maxima in the temporal Raman response (included in orange). We note that this trajectory corresponds to a broad set of sweeping cavity soliton solutions, and various forms of inter-solitons motion may be generated from the interplay of modelocking conditions and Raman coupling strength. A different experimental example for increased coupling strength is shown in Fig.4a, revealing meta-stable binding over many cavity round-trips at certain beating maxima of the lattice motion (maxima indicated as grey lines). More examples of sweeping trajectories from a prism-based oscillator are found in the Supplementary Information.

Finally, we want to emphasize key differences of the presented mechanism compared to soliton interactions in related nonlinear systems: a) Systems based on bulk glasses or fibers: In the absence of long-range translational symmetry, Raman selection rules for crystalline optical phonons no longer apply and the phonon density of states of the entire Brillouin zone contribute to a rapid quenching of the temporal Raman response, well known from silica fiber[25]. Impulsive stimulation of low-frequency acoustic waves, on the other hand, couples temporal solitons after reflection of sound waves[39–41]. b) Passive propagation: Strong Raman interactions induce continuous Raman scattering and self-frequency shifts contrary to active cavities[42], effectively suppressing spectral coherence[43]. Moreover, the enhancement of phase modulations via the Kerr-lens mechanism and amplification is absent.

In our experimental configuration, the cavity roundtrip time exceeds the phonon lifetimes, however, the presented coupling mechanism might be applied to long-lived optical phonons and miniaturized cavities. Active crystalline micro-resonators may offer further degrees to tailor and strengthen soliton-to-phonon coupling[9,10]. Beyond phonon stimulation via a single soliton and the coherent control of phonon dynamics via tailored soliton sequences[4,44,6,45] (Fig. 4b, left and middle), the feedback in a cavity with an optical mode spacing corresponding to phonon modes may enable resonant phonon pumping, stabilization of Terahertz pulse trains or "soliton crystals"[46], and eventually allow for ultrafast hybrid photonic-phononic devices and phonon-lasing[11,47], illustrated in Figure 4b (right).

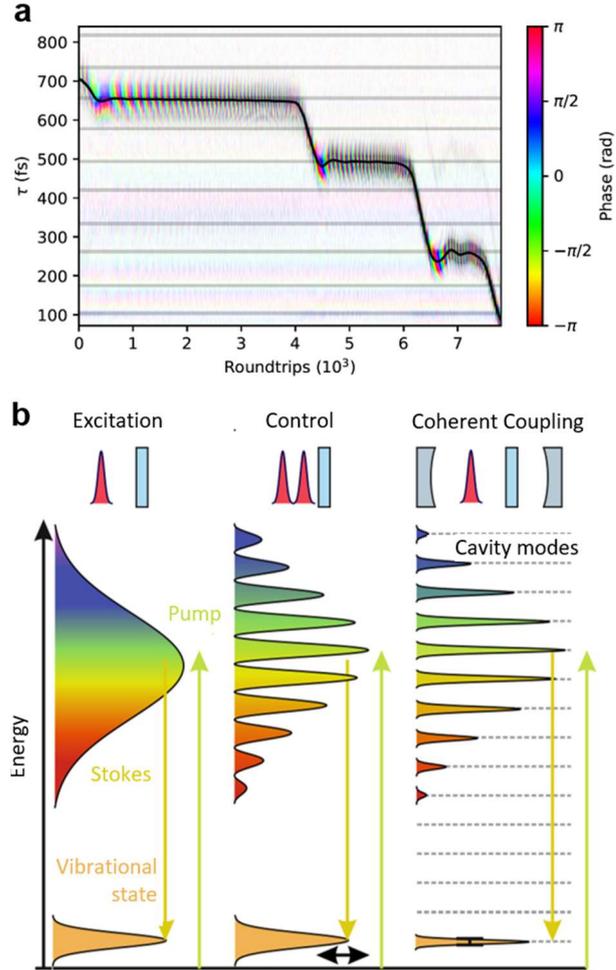

Figure 4. Coherent coupling of photonic to phononic wave-packets: a) Depending on coupling strength, stable or meta-stable soliton bound-states are generated at beating maxima of the coherent Raman waveform. Experimental data in according to the representation in Fig.2a are shown at increased effective binding along with the theoretical Raman maxima (indicated in grey). b) Left: Optical quanta of a single pulse excite vibrations for difference frequencies at the phonon energy. Middle: Control of phonon motion via a soliton sequence and coupling of solitons, vice versa. Right: Inside a cavity, the optical mode-spacing may match the phonon frequency, allowing for feedback and coherent coupling of soliton wave-packets to phonon motion. Resonant pumping results, e.g., in narrowing of phonon linewidths.

In conclusion, we reveal impulsive stimulated Raman scattering of optical phonons as a fundamental driving force for the coupling of ultrafast temporal cavity solitons in crystalline active media. The coupling is mediated via the Kerr nonlinearity in temporal and spatial domains, and manifests upon inter-soliton motion. The impulsively stimulated Raman waves generate stable and meta-stable soliton states at Terahertz phonon quantum beats, with more complex trajectories accessible via cavity tuning. Our results might enable the exploration of transient cavity soliton dynamics for rapid and autonomous, coherent time-domain sampling of optical phonons. The underlying coupling mechanism between temporal solitons and phonons offers unprecedented routes to ultrafast cavity phononics[12,48].

**Methods Summary:**

The experiments are based on a commercial 20-fs Kerr-lens mode-locked Ti:sapphire laser oscillator. High-speed soliton detection is implemented via two channels of a real-time oscilloscope with 40 GSa/s sampling rate and 8 GHz analog bandwidth, using direct detection by a fast photodiode and spectral interferometry via the time-stretch dispersive Fourier transformation (TS-DFT), respectively. Spectra are dispersed using 600m length of single mode optical fiber, enabling the resolution of soliton separations up to 2000 fs. Details on data analysis and simulation are provided in the Supplementary Information.

**Acknowledgements:**

We thank D. Brüggemann and M. Lippitz for technical support, and M.V. Axt and D.R. Solli for valuable discussions.

**Author contributions:**

All authors contributed to the experiment, data analysis and writing of the manuscript.